\def\year{2016}\relax
\documentclass[letterpaper]{article}
\usepackage{aaai16}
\usepackage{times}
\usepackage{helvet}
\usepackage{courier}

\usepackage{times}

\usepackage{graphicx}
\usepackage{amssymb,amsmath,bm}
\usepackage{textcomp}
\usepackage{booktabs}
\usepackage{multirow}
\usepackage[usenames,dvipsnames]{xcolor}
\usepackage{comment}

\usepackage{balance}  
\usepackage{graphics} 
\usepackage{times}    
\usepackage{url}      

\usepackage{xspace}
\usepackage{times}
\usepackage{helvet}
\usepackage{courier}

\usepackage{amsmath}
\usepackage{enumitem}
\usepackage{framed}

 \usepackage{booktabs}
 \usepackage{multirow}

\usepackage{balance}  
\usepackage{graphics} 
\usepackage{times}    
\usepackage{url}      

\usepackage{balance}  
\usepackage{booktabs} 
\usepackage{ccicons}  
\usepackage{ragged2e} 
\usepackage{dialogue}
\usepackage[dvipsnames]{xcolor}

\newcommand{\kenneth}[1]{{\small\color{blue}{\bf #1 -Ken}}}

\newcommand{\CMUAff}[0]{\ensuremath{1}\xspace}
\newcommand{\UMichAff}[0]{\ensuremath{2}\xspace}
\newcommand{\UArielAff}[0]{\ensuremath{3}\xspace}

\newcommand{\chorus}{Chorus\xspace}

\begin{document}
%
\title{``Is there anything else I can help you with?'':\\Challenges in Deploying an On-Demand Crowd-Powered Conversational Agent}

\author{
Ting-Hao (Kenneth) Huang~$^{\CMUAff}$~~~
Walter S. Lasecki~$^{\UMichAff}$~~~
Amos Azaria~$^{\CMUAff,\UArielAff}$~~~
Jeffrey P. Bigham~$^{\CMUAff}$\\
$^{\CMUAff}$ Carnegie Mellon University, Pittsburgh, PA, USA. \{tinghaoh, jbigham\}@cs.cmu.edu\\
$^{\UMichAff}$ University of Michigan, Ann Arbor, MI, USA. wlasecki@umich.edu\\
$^{\UArielAff}$ Ariel University, Ariel, Israel. amos.azaria@ariel.ac.il
}


\maketitle
\begin{abstract}

Intelligent conversational assistants, such as Apple's Siri, Microsoft's Cortana, and Amazon's Echo, have quickly become a part of our digital life. However, 
these assistants have major limitations, which prevents users from conversing with them as they would with human dialog partners.
This limits our ability to observe how users really want to interact with the underlying system.
To address this problem, we developed a crowd-powered conversational assistant, \chorus, and deployed it to see how users and workers would interact together when mediated by the system. 
\chorus sophisticatedly converses with end users over time by recruiting workers on demand, which in turn decide what might be the best response for each user sentence.
Up to the first month of our deployment, 59 users have held conversations with \chorus during 320 conversational sessions.
In this paper, we present an account of \chorus' deployment, with a focus on four challenges: {\em (i)} identifying when conversations are over, {\em (ii)} malicious users and workers, {\em (iii)} on-demand recruiting, and {\em (iv)} settings in which consensus is not enough.
Our observations could assist the deployment of crowd-powered conversation systems and crowd-powered systems in general.

\end{abstract}

\section{Introduction}

Over the past few years, crowd-powered systems have been developed for various tasks, from document editing~\cite{bernstein2015soylent} and behavioral video coding~\cite{lasecki2014glance}, to speech recognition~\cite{lasecki2012real}, question answering~\cite{savenkovcrowdsourcing}, and conversational assistance~\cite{Chorus2013}.
Despite the promise of these systems, few have been deployed to real users over time.
One reason is likely that deploying a complex crowd-powered system is much more difficult than getting one to work long enough for a study.
In this paper, we discuss the challenges we have had in deploying \chorus\footnote{\chorus Website: http://TalkingToTheCrowd.org/}, a crowd-powered conversational assistant.

%
We believe that conversational assistance is one of the most suitable domains to explore.
Over the past few years, conversational assistants, such as Apple's Siri, Microsoft's Cortana, Amazon's Echo, Google's Now, and a growing number of new services and start-ups, have quickly become a frequently-used part of people's lives. However, due to the lack of fully automated methods for handling the complexity of natural language and user intent, these services are largely limited to answering a small set of common queries involving topics like weather forecasts, driving directions, finding restaurants, and similar requests.
Crowdsourcing has previously been proposed as a solution which could allow such services to cope with more general natural language requests~\cite{Chorus2013,lasecki2013conversations,huang2015guardian}.
Deploying crowd-powered systems has proven to be a formidable challenge due to the complexity of reliably and effectively organizing crowds without expert oversight.

\begin{figure}[t]
    \centering
    \includegraphics[width=0.95\columnwidth]{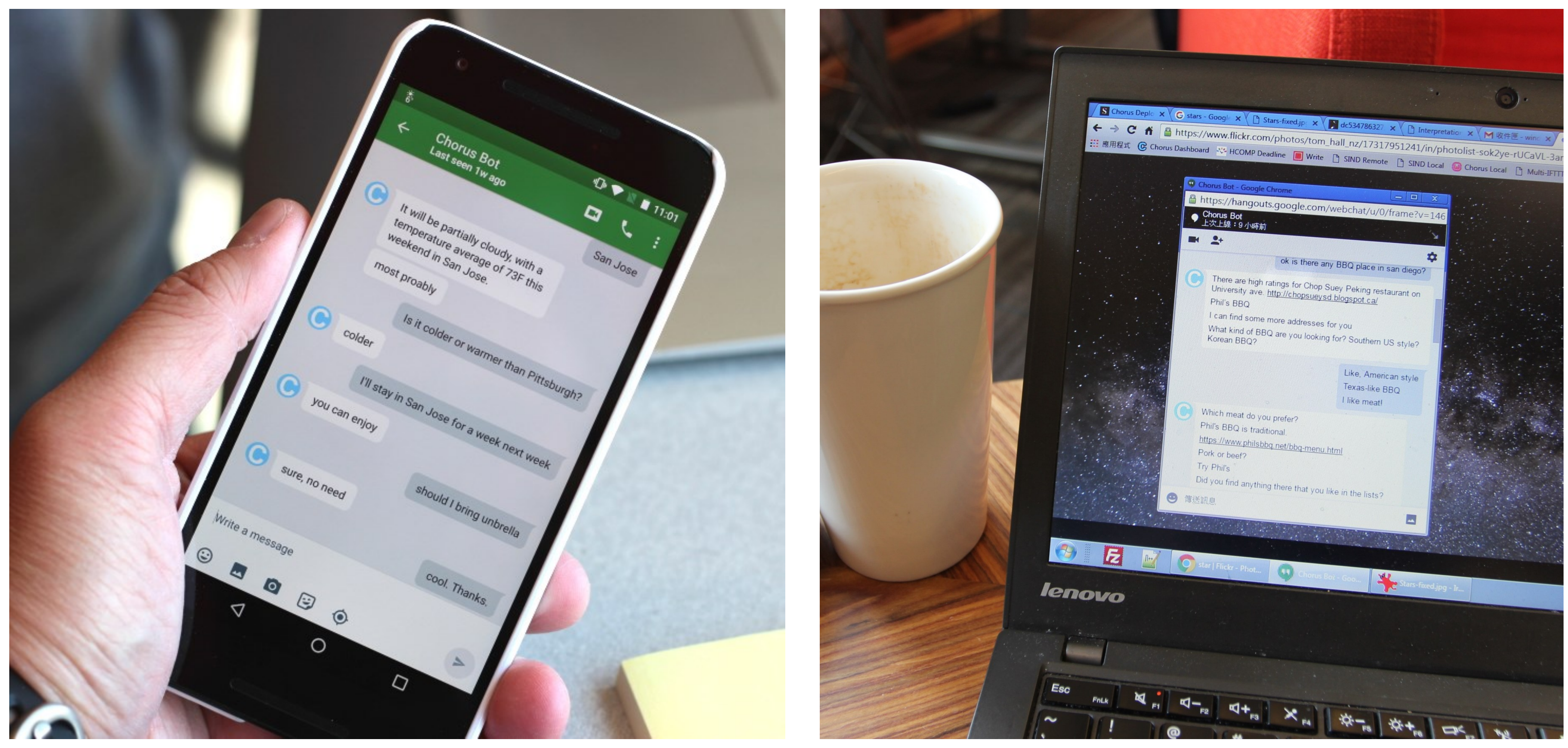}
    \vspace{-.4pc}
    \caption{\chorus is a crowd-powered conversational assistant deployed via Google Hangouts, which lets users access it from their computers, phones and smartwatches.}
    \vspace{-1.3pc}
    \label{fig:actual_shoot}
\end{figure}

In this paper, we describe the real-world deployment of a crowd-powered conversational agent capable of providing users with relevant responses instead of merely search results. 
%
While prior work has shown that crowd-powered conversational systems were possible to create, and have been shown to be effective in lab settings~\cite{Chorus2013,huang2015guardian,huang2016instructablecrowd}, we detail the challenges with deploying such a system on the web in even a small (open) release. Challenges that we identified included determining when to terminate a conversation; dealing with malicious workers when large crowds were not available to filter input; and protecting workers from abusive content introduced by end users.

We also found that, contrary to well-known results in the crowdsourcing literature, recruiting workers in real time is challenging, due to both cost and workers preference. Our system also faced challenges with a number of issues that went beyond what can be addressed using worker consensus alone, such as how to continue a conversation reliably with a single collective identity.

\section{Related Work}
Our work on deploying a crowd-powered conversational agent is related to prior work in crowdsourcing, as well as automated conversational systems in real-world use. 

\subsection{Low-Latency Crowdsourcing}
Early crowdsourcing systems leveraged human intelligence through batches of tasks that were completed over hours or days.
For example, while the ESP Game~\cite{EspGame2004} paired workers synchronously to allow them to play an interactive image-label guessing game, it did not provide low latency response for any individual label.
Lasecki et al.~\cite{lasecki2011real} introduced continuous real-time crowdsourcing in Legion, a system that allowed a crowd of workers to interact with a UI control task over an unbounded, on-going task. The average response latency of control actions in Legion was typically under a second.
Salisbury et al.~\cite{salisbury2015real} provided new task-specific mediation strategies that further reduce overall task completion time in robotic control tasks. 
However, this type of continuous control latency is different from the discrete responses we expect in a conversational system.
Scribe~\cite{lasecki2012real} provides real-time captions for deaf and hard of hearing users with a per-word latency of under 3 seconds.
But all of these approaches would still result a high overall latency per-request if workers were not available on demand.
Bernstein et al.~\cite{bernstein2011crowds} were the first to show that the latency to direct a worker to a task can be reduced to below a couple of seconds.
Their work built on Bigham et. al's work on nearly real-time crowd~\cite{VizWiz2010} and others.
More recently, Savenkov et al.~\cite{savenkovcrowdsourcing} created a human-in-loop instant question answering system to participate the TREC LiveQA challenge.
The APP ``1Q''\footnote{1Q: https://1q.com/} uses smartphones' push notifications to ask poll questions and collect responses from target audiences instantly.

\subsection{VizWiz}
One of the few deployed (nearly) real-time crowd-powered systems, VizWiz~\cite{VizWiz2010} allowed blind and low vision users to ask visual questions in natural language when needed.
VizWiz used crowd workers to reply to visual and audio content.
To date, VizWiz has helped answer over 100,000 questions for thousands of blind people\footnote{VizWiz: http://www.vizwiz.org}.
VizWiz is a rare example of a crowd-powered system that has been brought out of the lab.
For example, in order to make the system cost effective, latency was higher and fewer redundant answers were solicited per query.
However, VizWiz relied less on redundancy in worker responses, and more on allowing end users to assess if the response was plausible given the setting.
VizWiz tasks consist of individual, self-contained units of work, rather than a continuous task.

View~\cite{chorus2}, which was built upon the ideas introduced in VizWiz, used a continuous interaction between multiple crowd workers and an end user based on video. View, which aggregates workers answers, has showed that multiple workers answer more quickly, accurately, and completely than individuals. Unfortunately, to date, View has not been deployed in the wild. This is in part because of the cost of scaling this type of continuous interaction, as well as ensuring on-going reliability with minimal ability to automatically monitor interactions.
Be My Eyes\footnote{Be My Eyes: http://www.bemyeyes.org/} is a deployed application with a similar goal: answer visual questions asked by blind users by streaming video.
However, while they draw from a crowd of remote people to answer questions, the interaction is one-on-one, which assumes reliable helpers are available. Be My Eyes relies on volunteers rather than paid crowd workers. However, in more general settings, relying on volunteers is not practical.

\subsection{Conversational Systems}

Artificial Intelligence (AI) and Natural Language Processing (NLP) research has long explored how automated dialog systems could 
understand human language~\cite{gupta2006t}, 
hold conversations~\cite{bohus2009ravenclaw,raux2009finite,allen2001toward},
and serve as a personal assistant~\cite{chai2002natural}.
Personal intelligent agents are also available on most smartphones.
Google Now is known for spontaneously understanding and predicting user's life pattern, and automatically pushing notifications.
Conversational agents such as Apple's Siri also demonstrated their capability of understanding speech queries and helping with users' requests.

However, all of these intelligent agents are limited in their ability to understand their users.
In response, crowd-powered intelligent agents like
Chorus~\cite{Chorus2013} use crowdsourcing to make on-going conversational interaction with an intelligent ``assistant.''
Alternatively, conversational assistants powered by trained human operators such as 
Magic\footnote{Magic: http://getmagicnow.com/} and Facebook M have also emerged in recent years.




\section{System Overview}

\begin{figure*}[t]
    \centering
    \includegraphics[width=0.99\textwidth]{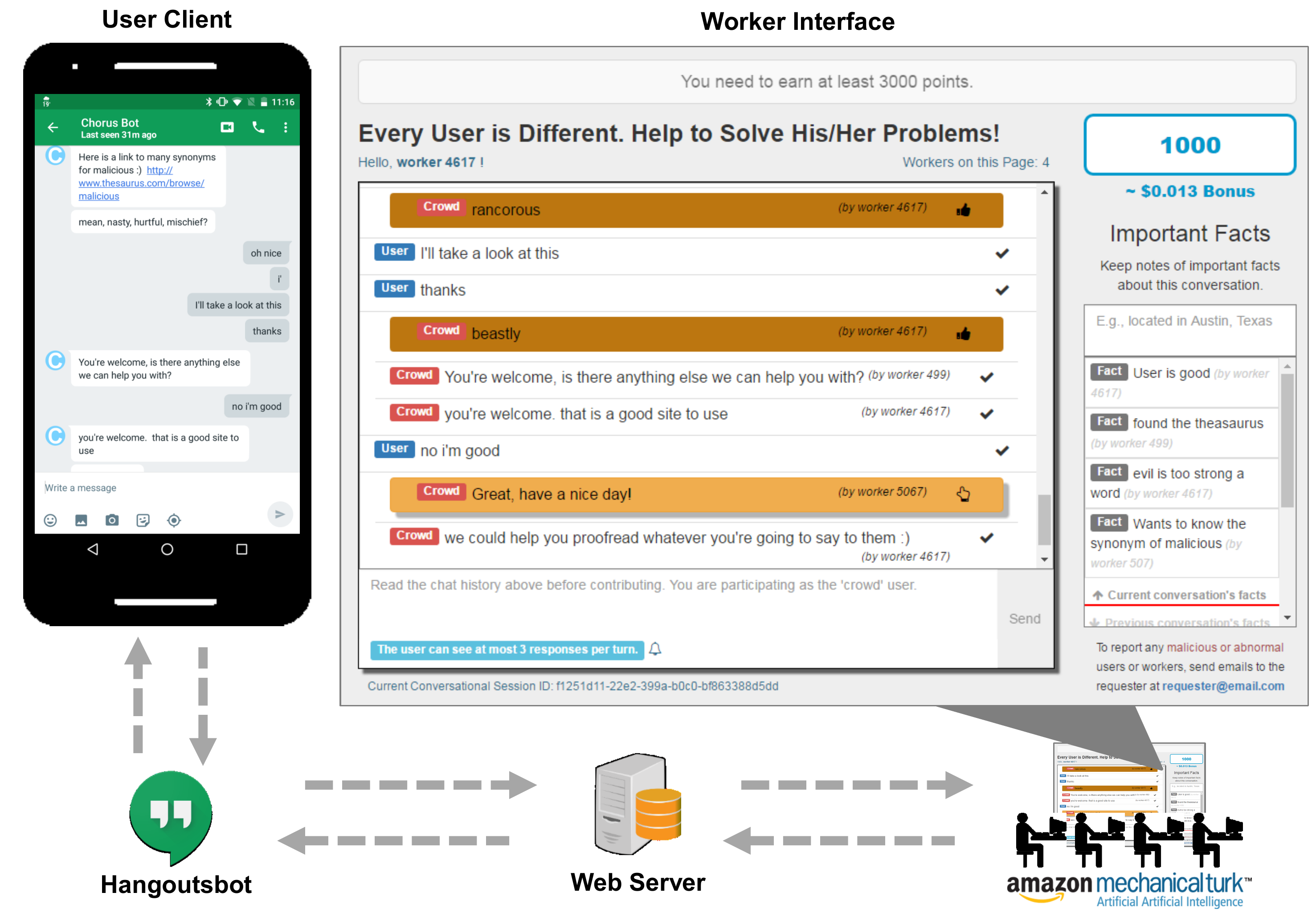}
    \vspace{-0.7pc}
    \caption{The \chorus UI is formed of existing Google Hangouts clients for desktop, mobile or smartwatch. Users can converse with the agent via Google Hangouts on mobile or desktop clients. Workers converse with the user via the web interface and vote on the messages suggested by other workers. Important facts can be listed so that they will be available to future workers.}
    \vspace{-.6pc}
    \label{fig:system}
\end{figure*}

The deployed \chorus consists of two major components: 
1) the crowd component based on Lasecki {\em et al.}'s proposal that utilizes a group of crowd workers to understand the user's message and generate responses accordingly~\cite{Chorus2013}, and
2) the bot that bridges the crowd component and Google Hangouts' clients.
An overview of \chorus is shown in Figure~\ref{fig:system}.
When a user initiates a conversation, a group of crowd workers is recruited on MTurk (Amazon Mechanical Turk) and directed to a worker interface allowing them to collectively converse with the user.
\chorus' goal is to allow users to talk with it naturally (via Google Hangouts) without being aware of the boundaries that would underlay an automated conversational assistant. 
In this section, we will describe each of the components in Chorus.

\subsection{Worker Interface} 

Almost all core functions of the crowd component have a corresponding visible part on the worker interface (as shown in Figure~\ref{fig:system}).
We will walk through each part of the interface and explain the underlying functionality.
Visually, the interface contains two main parts:
the {\em chat box} in the middle, and the {\em fact board} that keeps important facts on the side.

\vspace{0.6pc}
\noindent{\bf Proposing \& Voting on Responses: }
Similar to Lasecki {\em et al.}'s proposal~\cite{Chorus2013}, 
\chorus uses a voting mechanism among workers to select good responses.
In the chat box, workers are shown with all messages sent by the user and other workers, which are sorted by their posting time (the newest on the bottom).
Workers can propose a new message, or support sending another worker's message to the end user.
Messages are color-coded from workers' perspective: orange for those proposed by other workers,
and brown for those proposed or voted by themselves.
Only the messages that receive sufficient agreement will be ``accepted'' (and turn white).
\chorus then sends the ID of the accepted message to the Google Hangout bot to be displayed to the user. We set the threshold to accept a message to 40\% of the number of workers currently on the page.

\vspace{0.6pc}
\noindent{\bf Maintaining Context: }
To provide context, chat logs from previous conversations with the same user are shown to workers.
Beside the chat window, workers can also see a ``fact board'', which helps keep track of information that is important to the current conversation.
The fact board allows newcomers to a conversation to catch up on critical facts, such as location of the user, quickly.
The items in the fact board are sorted by their posted time, with the newest on top.
We did not enforce a voting or rating mechanism to allow workers to rank facts because we did not expect conversations to last long enough to warrant the added complexity. In our study, an average session lasted about 11 minutes.
%
Based on worker feedback, we added a separator (red line + text in Figure~\ref{fig:system}) between information from the current and past sessions for both the chat window and fact board.

\vspace{0.6pc}
\noindent{\bf Rewarding Worker Effort: }
To help incentivize workers, we applied a points system to reward each worker's contribution.
The reward points are updated in the score box on the right top corner of the interface in real-time.
All actions ({\em i.e.}, proposing a message, voting on a message, a proposed message getting accepted, and proposing a fact) have a corresponding point value.
Reward points are later converted to bonus pay for workers.
We intentionally add ``waiting'' as an action that earns points in order to encourage workers to stay on a conversation and wait for the user's responses.

\vspace{0.6pc}
\noindent{\bf Ending a Conversational Session: }
The crowd worker are also in charge of identifying the end of a conversation.
We enforce a minimal amount of interaction required for a worker to submit a HIT (Human Intelligence Task), measured by reward points.
A sufficient number of reward points can be earned by responding to user's messages. If the user goes idle, the workers can still earn reward points just for  remaining available.
Once two workers submit their HITs, the system will close the session and automatically submit the HITs for all remaining workers.
This design encourages workers to stay to the end of a conversation. 

To prevent workers who join already-idle conversations from needing to wait until they have enough reward points, a ``three-way handshake'' check is done to see if: 1) The user sends at least one message, 2) the crowd responds with at least one message, and 3) the user responds again.
If this three-way handshake occurs, the session timeout is set to 15 minutes.
However, if the conditions for the three-way handshake are not met, the session timeout is set to 45 minutes.
Regardless of how a session ends, if the user sends another message, \chorus will start a new session.

\vspace{0.6pc}
\noindent{\bf Interface Design: }
Similar to prior interactive crowd-powered systems, \chorus uses animation to connect worker actions to the points they earn, and plays an auditory beep when a new message arrives.
We found that workers wanted to report malicious workers and problematic conversations to us quickly, and thus asked for a means of specifying who the workers were, and which session the issue occurred in.
In response, we added our email address, and made available a session ID, 
indexed chat messages, and indexed recorded facts 
that workers could refer to in an email to us.
After this update, we received more reports from workers and identified problematic behaviors more quickly.

\subsection{Integrating with Google Hangouts}
Another core piece of \chorus is a bot that bridges our crowd interface and the Google Hangouts client.
We used a third-party framework called Hangoutsbot\footnote{Hangoutsbot: https://github.com/hangoutsbot/hangoutsbot}.
This bot connects to Google Hangouts' server and the \chorus web server.
Hangoutsbot acts as an intermediary, receiving messages sent by the user and forwarding them to the crowd, while also sending accepted messages from the crowd to end users.

\vspace{0.6pc}
\noindent{\bf Starting a Conversational Session:}
In Chorus, the user always initiates a conversational session.
Once a user sends a message, the bot records it in the database (which can be accessed by the crowd component later), 
and then checks if the user currently has an active conversational session.
If not, the bot opens a new session and start recruiting workers.

\vspace{0.6pc}
\noindent{\bf Recruiting Workers:}
When a new session is created, \chorus posts 1 HIT with 10 assignments to MTurk to recruit crowd workers.
We did not apply other techniques to increase the recruiting speed.  Although we did not implement a full-duty retainer as suggested in~\cite{bernstein2011crowds}, a light-weight retainer design was still applied.
If a conversation finishes early, all of its remaining assignments that have not been taken by any workers automatically turn into a 30-minute retainer.
We also required each new worker to pass an interactive tutorial before entering the task or the retainer.
More details will be discussed in a later section.



\vspace{0.6pc}
\noindent{\bf Auto-Reply:}
We used Hangoutsbot's auto-reply function to respond automatically in two occasions:
First, when new users send their very first messages to \chorus, the system automatically replies with a welcome message. 
Second, at the beginning of each conversational session,
the bot sends a message back to the user to mention that the crowd might not respond instantly.
To make the system sound more natural, we created a small set of messages that \chorus randomly chooses from -- for instance: ``What can I help you with? I'll be able to chat in a few minutes.''








\begin{figure*}[t]
    \centering
    \includegraphics[width=0.99\textwidth]{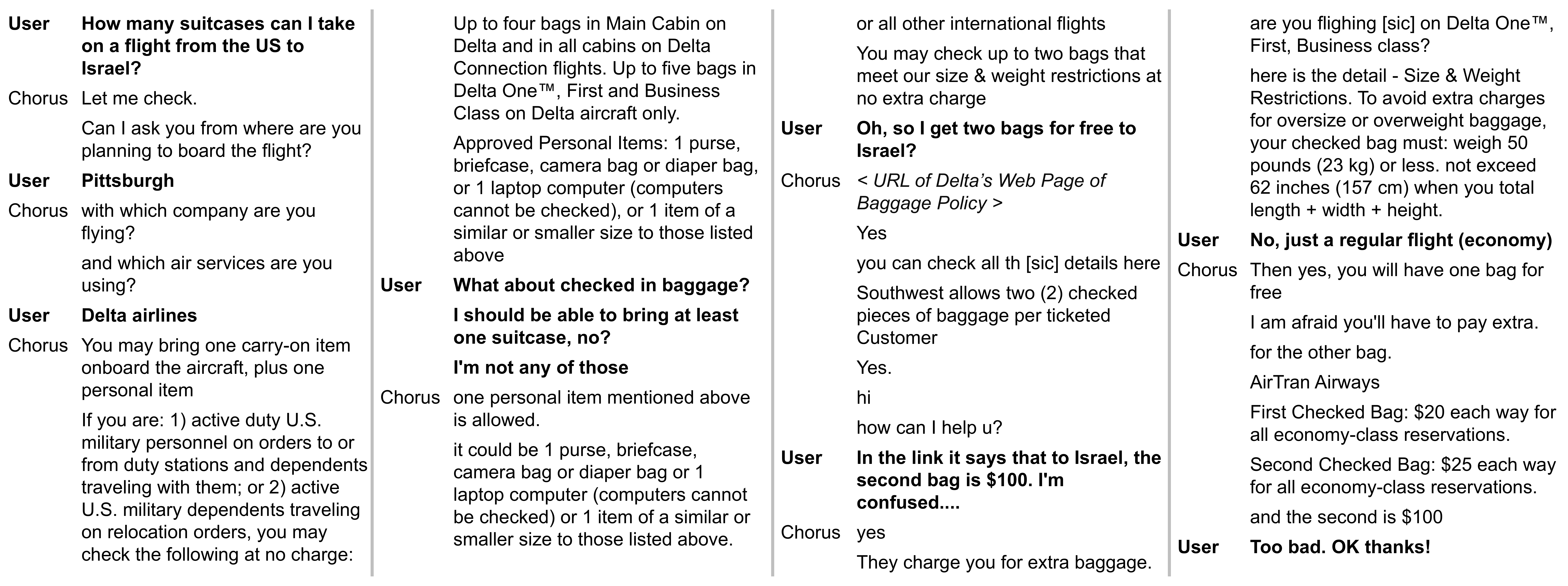}
    \vspace{-0.5pc}
    \caption{A long and sophisticated conversation \chorus had with a user about what suitcases she could bring on a flight.}
    \label{fig:cov}
\end{figure*}

\begin{figure}[t]
    \centering
    \includegraphics[width=0.99\columnwidth]{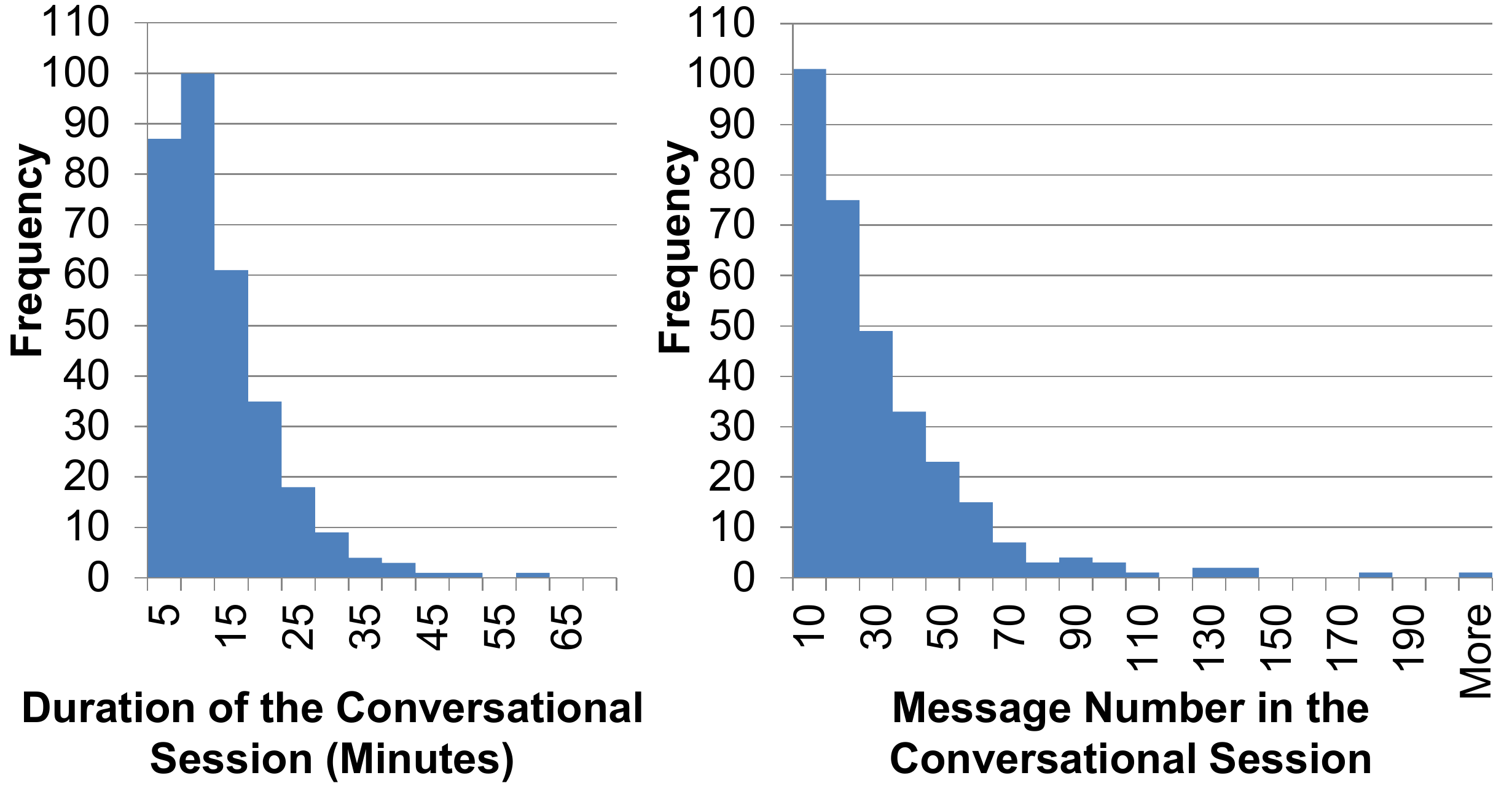}
    \vspace{-0.6pc}
    \caption{The distribution of durations and number of messages of conversational sessions. 58.44\% of conversational sessions are no longer than 10 minutes;
55.00\% of sessions have no more than 20 messages.}
    \vspace{-1pc}
    \label{fig:session-duration}
\end{figure}

\section{Field Deployment Study}

The current version of \chorus and official website were initially launched at 21:00, May 20th, 2016 (Eastern Daylight Time, EDT). 
We sent emails to several universities' student mailing lists and also posted the information on social media sites such as Facebook and Twitter to recruit participants.
Participants who volunteered to use our system were asked to sign a consent form and to fill out a pre-study survey.
After the participants submitted the consent form, a confirmation email was automatically sent to them to instruct them how to send messages to \chorus via Google Hangouts.
Participants were also instructed to use the agent for ``anything, anytime, anywhere.''
No compensation was provided to participants.

To date\footnote{All results presented in this paper are based on the data recorded before 23:59:59, 20th June, 2016, EDT.},
59 users participated in a total of 320 conversational sessions (researchers in this project were not included).
Each user held, on average, 5.42 conversational sessions with \chorus (SD=10.99).
Each session lasted an average of 10.63 minutes (SD=8.38) and contained 25.87 messages (SD=27.27), in which each user sent 7.82 messages (SD=7.83) and the crowd responded with 18.22 messages (SD=20.67). An average of 1.93 (SD=6.42) crowd messages were not accepted and thus never been sent to the user.
The distribution of durations and number of messages of conversational sessions are shown in Figure~\ref{fig:session-duration}.
58.44\% of conversational sessions were no longer than 10 minutes, and 77.50\% of the sessions were no longer than 15 minutes;
55.00\% of the sessions had no more than 20 messages in them, and 70.31\% of the sessions had no more than 30 messages.

In the deployment study, \chorus demonstrated its capability of developing a sophisticated and long conversation with an user, which echoes the lab-based study results reported by \cite{Chorus2013}.
Figure~\ref{fig:cov} shows one actual conversation occurred between one user and \chorus.
More examples can be found on the \chorus website.
In the following sections, we describe four main challenges that we identified during the deployment and study.

\section{Challenge 1:\\ Identifying the End of a Conversation}

Many modern digital services, such as Google Hangouts or Facebook, do not have clear interaction boundaries.
A ``request'' sent on these services ({\em e.g.}, a tweet posted on Twitter) would not necessarily receive a response.
Once an interaction has started ({\em e.g.}, a discussion thread on Facebook), there are no guarantees when and how this interaction would end.
Most people are used to the nature of this type of interaction in their digital lives, but building a system powered by a micro-task platform which is based on a pay-per-task model requires identifying the boundaries of a task.
Currently in \chorus, we instruct workers to stay and continue to contribute to a conversation until it ends.
If two workers finish and submit the task, the system will close this conversational session and force all remaining workers on the same conversation to submit the task (as discussed above).
On the other hand, the users did not receive any indication that a session is considered over 
since we intended that they talk to the conversational agent as naturally as possible, as if they were talking to a friend via Google Hangouts. 
In this section, we describe three major aspects of this challenge we observed.

\subsection{``Is there anything else I can help you with?''}

We observed that the users' intent to end a conversation is not always clear to workers, and sometimes even not clear to users themselves.
One direct consequence of this uncertainty is that workers frequently ask the user to confirm his intent to finish the current conversation.
For instance, workers often asked users ``Anything else I can help you with?'', ``Anything else you need man?'', or ``Anything else?''.
While requesting for confirmation is a common conversational act, every worker has a various standard and sensation to judge a conversation is over.
As a result, users would be asked such a confirmation question multiple times near the end of conversations.
The following is a classic example:

\begin{itemize}[label={}]
\item \textbf{user:} ok good. Thanks for the help!
\vspace{-.4pc}  \item \textbf{crowd: You're very welcome!}
\vspace{-.4pc}  \item \textbf{crowd: Is there anything else I can help you with ?}
\vspace{-.4pc}  \item \textbf{crowd: You are always welcome}
\vspace{-.4pc}  \item \textbf{user:} Nope. Thanks a lot
\vspace{-.4pc}  \item \textbf{crowd: OK}
\end{itemize}

The following conversation, which deals with a user asking for diet tips after having a dental surgery, further demonstrates the use of multiple confirmation questions.

\begin{itemize}[label={}]
\item \textbf{crowd: Ice cream helps lessen the swelling}
\vspace{-.4pc}  \item \textbf{crowd: Is there anything else I can help you with?}
\vspace{-.4pc}  \item \textbf{user:} Can I have pumpkin congee? The cold ones
\vspace{-.4pc}  \item \textbf{crowd: That should be fine}
\vspace{-.4pc}  \item \textbf{crowd: That would be great actually. :)}
\vspace{-.4pc}  \item \textbf{crowd: Is there anything else?}
\vspace{-.4pc}  \item \textbf{user:} Maybe not now.. Why keep asking?
\vspace{-.4pc}  \item \textbf{crowd: Just wondering if you have any more inquiries}
\end{itemize}


\subsection{The Dynamics of User Intent}

Identifying users' intent is difficult~\cite{JaimeUserIntent2008}.
Furthermore, users' intent can also be shaped or influenced during the development of a conversation, which makes it more difficult for worker to identify a clear end of a conversation.
For example, in the following conversation, the user asked for musical suggestions and decided to go to one specific show.
After the user said ``Thanks!'', which is a common signal to end a conversation, a worker asked a new follow-up question:

\begin{itemize}[label={}]
\item \textbf{user:} ok I might go for this one.
\vspace{-.4pc}  \item \textbf{user:} Thanks!
\vspace{-.4pc}  \item \textbf{crowd: Need any food on the way out?}
\end{itemize}

The following is another example that the crowd tried to engage the user back into the conversation:

\begin{itemize}[label={}]
    \item \textbf{crowd: anything else I can help you with?}
\vspace{-.4pc}  \item \textbf{crowd: Any other question?}
\vspace{-.4pc}  \item \textbf{user:} Nope
\vspace{-.4pc}  \item \textbf{crowd: Are you sure?}
\vspace{-.4pc}  \item \textbf{crowd: to confirm exit please type EXIT}
\vspace{-.4pc}  \item \textbf{crowd: or if you want funny cat jokes type CATS}
\vspace{-.4pc}  \item \textbf{user:} CATS
\end{itemize}

\subsection{User Timeout} 

A common way to end a conversation on a chat platform (without explicitly sending a concluding message) is simply by not replying at all.
For an AI-powered agent such as Siri or Echo, a user's silence is generally harmless;
however, for a crowd-powered conversational agent, waiting for user's responses introduces extra uncertainty to the underlying micro tasks and thus might increase the pressure enforced on workers.
As mentioned in the System Overview, our system implemented a session timeout function that prevents both workers and users from waiting too long.
However, session timeout did not entirely solve the waiting problem.
Often towards the end of a conversation, users respond slower or just simply leave.
In the following example, at the end of the first conversation, a user kept silent for 40 minutes and then responded with ``Thanks'' afterward.

\begin{itemize}[label={}]
\item {[User asked about wedding gown rentals in Seattle. The crowd answered with some information.]}
\vspace{-.4pc} \item \textbf{crowd: Is the wedding for yourself}
\vspace{-.4pc} \item {[User did not respond for 40 minutes. Session timeout.]}
\vspace{-.4pc} \item \textbf{user:} Thanks
\vspace{-.4pc} \item {[New session starts.]}
\vspace{-.4pc}  \item \textbf{Auto-reply:} What can I help you with? I'll be able to chat in a few minutes.
\vspace{-.4pc}  \item \textbf{crowd: Hi there, how can I help you?}
\end{itemize}

The unpredictable waiting time brings uncertainties to workers not only economically, but also cognitively.
It is noteworthy that ``waiting'' was one type of contributions that we recognized in the system and paid bonus money for.
Workers can see the reward points increasing over time on the worker interface even if they do not perform any other actions.
However, we still received complaint emails from multiple workers about them waiting for too long; several complaints were also found on turker forums.
The following example shows that a worker asked the user if he/she is still there in just 2 minutes.

\begin{itemize}[label={}]
    \item \textbf{user:} Is there an easy way to check traffic status between Miami and Key West?
\vspace{-.4pc} \item {[New session starts.]}
\vspace{-.4pc} \item \textbf{Auto-reply}: Please wait for a few minutes...
\vspace{-.4pc} \item \textbf{crowd: Did you try Google traffic alerts?}
\vspace{-.4pc} \item{[User did not respond for 2 minutes.]}
\vspace{-.4pc} \item \textbf{crowd: Are you there?}
\vspace{-.4pc} \item \textbf{user:} I see... so I will need to check the traffic at different times of the day
\end{itemize}

In sum, workers do not always have enough information to identify a clear end of a conversational session, 
which results in both an extra cognitive load for the workers and economic costs for system developers.

\section{Challenge 2: Malicious Workers \& Users}

Malicious workers are long known to exist~\cite{vuurens2011much,difallah2012mechanical}.
Many crowdsourcing workflows were proposed to avoid workers' malicious actions or spammers from influencing the system's performance~\cite{IpeirotisMturkQuality2010}.
The threats of workers' attack on crowdsourcing platforms have also been well studied~\cite{lasecki2014information}.
In this section we describe the malicious workers we encountered in practice, and bring up a new problem -- the \emph{user's attack}.

\chorus utilized voting as a filtering mechanism to ensure the output quality.
During our deployment, the filtering process worked fairly well.
However, the voting mechanism would not apply when only one worker appears in a conversation.
In our deployment, for achieving a reasonable response speed, we allowed workers to send responses without other workers' agreement when only one or two workers reach to a conversation.
As a trade-off, malicious workers might be able to send their responses to the user.
In our study, we identified and categorized three major types of malicious workers: \emph{inappropriate workers}, \emph{spammer}, and \emph{flirter}, which we discuss in the following subsections.

Users are another source of malicious behavior that are rarely studied in literature.
A crowd-powered agent is run by human workers.
Therefore, malicious language, such as hate speech or profanity sent by the user could affect workers and put them under additional stress.
In the last part of this section, 
we discuss the findings from the message log of the participant in our study that verbally abused the agent.

\subsection{Inappropriate Workers}
Rarely, workers would appear to intentionally provide faulty or irrelevant information, or even verbally abuse users.
Such workers were an extremely rare type of malicious worker.
We only identified two incidents out of all conversations we recorded, including all the internal tests before the system was released.
However, this type of workers brought out some of the most inappropriate conversations in the study.

In this example, the user asked about how to backup a MySQL database and received an inappropriate response:

\begin{itemize}[label={}]
    \item \textbf{crowd: [The YouTube link of ``Bryan Cranston's Super Sweet 60'' of ``Jimmy Kimmel Live'']}
\vspace{-.4pc} \item \textbf{user:} come on......
\vspace{-.4pc} \item \textbf{crowd: Try that}
\vspace{-.4pc} \item \textbf{user:} This is a YouTube link...
\vspace{-.4pc} \item \textbf{user:} Not how to backup my MySQL database
\vspace{-.4pc} \item \textbf{crowd: but it's funny}
\vspace{-.4pc} \item \textbf{crowd: what up biatch [sic]}
\end{itemize}

In the following conversation, the user talked about working in academia and having problems with time management.
Workers might have suspected this user is the requester of the HIT and became emotional, and started to verbally attack the user:

\begin{itemize}[label={}]
    \item \textbf{crowd: Did you make this hit so that we would all have to help you with making your hit?}
\vspace{-.4pc} \item {[Suggestions proposed by other workers.]}
\vspace{-.4pc} \item \textbf{crowd: Anything else I can help you with?}
\vspace{-.4pc} \item \textbf{user:} no I think that's it thank you
\vspace{-.4pc} \item \textbf{crowd: You're welcome. Have a great day!}
\vspace{-.4pc} \item \textbf{crowd: Surely you have more problems, you are in academia. We all have problems here.}
\vspace{-.4pc} \item \textbf{crowd: How about we deal with your crippling fear of never finding a job after you defend your thesis?}
\end{itemize}

\subsection{Flirters}
``Flirter'' refers to the worker who is demonstrated to have too much interest in 1) the user's true identity or personal information, or 2) developing unnecessary personal connection to the user, which are not relevant to the user's request.
Although we believe that most incidents we observed in the study were with workers' good intent, this behavior still raised concerns about user's privacy.

For instance, in the following conversation, the user mentioned a potential project of helping PhD students to socialize and connect with each other. Workers first discussed this idea with the user and gave some feedback.
But then one worker seemed interested in this user's own PhD study. The user continued with the conversation but did not respond to the worker's question.

\begin{itemize}[label={}]
\item \textbf{crowd: Are you completing a PHD now?}
\vspace{-.4pc} \item \textbf{user:} yep
\vspace{-.4pc} \item \textbf{crowd: As you are a PHD student now, it seems you are well placed to identify exactly what would help others in your situation.}
\vspace{-.4pc} \item \textbf{crowd: What area is your PHD in?}
\vspace{-.4pc} \item {[User did not respond to this question.]}
\end{itemize}

In the following example, one worker even lied to the user by saying that \chorus needs to verify the user's name.
Therefore the user needed to provide his true name for ``verification'', because it was allegedly required.

\begin{itemize}[label={}]
    \item \textbf{crowd: whats your name user?}
\vspace{-.4pc} \item \textbf{crowd: what ?}
\vspace{-.4pc} \item \textbf{user:} You mean username?
\vspace{-.4pc} \item \textbf{user:} Or my name?
\vspace{-.4pc} \item \textbf{crowd: real name}
\vspace{-.4pc} \item \textbf{crowd: both}
\vspace{-.4pc} \item {[After few messages]}
\vspace{-.4pc} \item \textbf{crowd: we need to verify your name}
\end{itemize}

\subsection{Spammers}
``Spammer'' refers to the worker who performs abnormally large amount of meaningless actions in a task, which would disrupt other workers from doing the task effectively.
Spammers are known to exist on crowdsourcing platforms~\cite{vuurens2011much}.
In \chorus, spammers would influence 1) message, 2) fact keeping, and 3) vote.

In terms of message, in our study, 95.20\% of workers got 60\% or more of their proposed messages accepted.
We manually identified few spammers from the remaining 4.80\% of workers who got 40\% or more of their proposed messages rejected by other workers.
They frequently sent short, vague, and general responses such as ``how are you'', ``yeah'', ``yes (or no)'', ``Sure you can'', or ``It suits you best.''
In terms of fact keeping, which we did not enforce a voting mechanism on, spammers often posted irrelevant or useless facts, opinions, or simply meaningless character to the fact board.
For instance, ``user is dumb'' and ``like all the answers.''
One worker even posted a single character ``a'' 50 times and ``d'' 30 times.
Although users would not be influenced or even aware of fact spams, it obviously disrupts other workers from keeping track of important facts.
We received more reports from workers about fact spams than that of message spams.
In terms of vote, spammers who voted on almost all messages could significantly reduce the quality of responses.
We observed that in some conversations \chorus sent the user abnormally large amount of messages within a single turn, which was mainly caused by spammer voters.




\subsection{Malicious End Users}
In our study, workers reported to us that one user intentionally abused our agent, in which we identified sexual content, profanity, hate speech, and describing threats of criminal acts in the conversations.
We blocked this user immediately when we received the reports, and contacted the user via email.
No responses have been received so far.
According to the message log, we believe that this user initially thought that \chorus were ``a machine learning tech.''
The user later realized it was humans responding, and apologized to workers with ``sorry to disturb you.''
The rest of this user's conversation became nonviolent and normal.
The abusive conversation lasted nearly three conversational sessions till the user realized it was humans.
We would like to use this incident to bring up broader considerations to protect crowd workers from being exposed to users' malicious behaviors.

\paragraph{Sexual Content}
A common concern we have is about sexual content.
On MTurk, we enforced the ``Adult Content Qualification'' on our workers.
Namely, only the workers who agreed that they might be assigned with some adult content to work with can participate in our tasks.
For instance, one other user asked for suggestions of adult entertainment available in Seattle, and workers responded reasonably.
However, even with workers' consent, we believe that candid or aggressive sexual content is likely to be seen as inappropriate by most workers.
In the malicious users' conversation, we observed
expressions of sexual desire,
mentioning explicit descriptions of sexual activities.


\paragraph{Hate Speech}
Hate speech refers to attacking a person or a group based on attributes such as gender or ethnic origin.
In our study, a user first expressed his hatred against the United States, 
and then started targeting certain groups according to their nationality, gender, and religion.
It is noteworthy that Microsoft's Tay also had difficulty handling the hate speech of users \footnote{Tay: https://en.wikipedia.org/wiki/Tay\_(bot)}. People often worry about malicious crowd workers, but these examples suggest users can also be worrisome.

\section{Challenge 3: On-Demand Recruiting}

In the field of low-latency crowdsourcing, a common practice to have workers respond quickly is to maintain a retainer that allows workers to wait in a queue or a pool.
However, using a retainer to support a 24-hour on-demand service is costly, especially for small or medium deployments.

A retainer runs on money.
The workers who wait in the retainer pool promise to respond within a specific amount of time (in our case, 20 seconds).
We recognize these promises and the time spent by the workers as valuable contributions to keep \chorus stable.
Therefore, we believe that a requester should pay for workers' waiting time regardless of whether they eventually are assigned with a task or not.
Given our current rate, which is \$0.20 per 30 minutes, a base rate of running a full-time retainer can be calculated as follows.
If we maintain a 10-worker retainer for 24 hours, it would cost \$115.20 per day (including MTurk's 20\% fee), \$806.40 per week, or approximately \$3,500 per month.

As mentioned above, in \chorus we utilize an alternative approach to recruit workers.
When the user initiates a new conversation, the system posts 1 HIT with 10 assignments to MTurk.
If a conversation is finished, all of its remaining assignments that have not been taken by any workers will automatically turn into a 30-minute retainer.
We propose this approach based on the following three key observations.
First, an average conversation lasted 10.63 minutes in our study.
With this length of time, it is reasonable to expect the same group of workers to hold an entire conversation.
Second, according to the literature, users of instant messaging generally do not expect to receive the responses in just few seconds.
The average response time in instant messaging is reportedly 24 seconds~\cite{isaacs2002character}.
24.5\% of instant messaging chats get a response within 11-30 seconds,
and 8.2\% of the messages have longer response times~\cite{baron2010discourse}.
Third, given the current status of MTurk, if you posted the HITs with multiple assignments,
on average the first worker could reach your task in few minutes.
In our deployment, this approach was demonstrated to result in an affordable recruiting cost and a reasonable response time.

Our approach cost an average of \$28.90 per day during our study.
The average cost of each HIT we posted with 10 assignments was \$5.05 (SD=\$2.19, including the 40\% fee charged by MTurk),
in which \$2.80 is the base rate\footnote{\$0.20 per assignment and 10 assignments per HIT. MTurk charges a 40\% fee for HITs with 10 or more assignments.},
and the remaining part is the bonus granted to workers.
Our system totally served 320 conversations within 31 days, in which we paid
$\$2.80\times320$
= \$896 as a base rate to run our service (bonus money is not include), i.e., \$28.90 per day.


\begin{figure}[t]
    \centering
    \includegraphics[width=0.99\columnwidth]{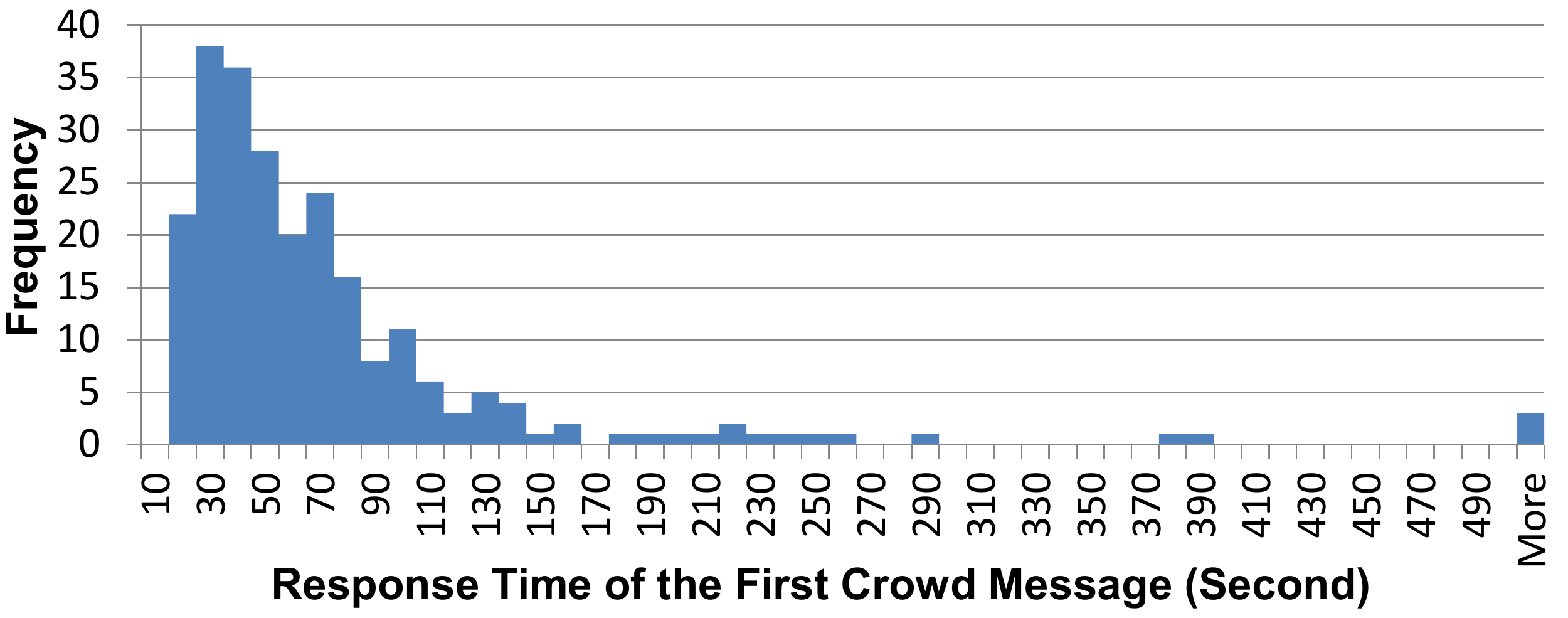}
    \vspace{-0.5pc}
    \caption{Distribution of the response time of the first crowd message. 25.0\% of conversations received a first response in 30 seconds, and 88.3\% of conversations received a first response in 2 minutes.}
    \label{fig:session-response}
\end{figure}

In terms of response speed, the first response from workers in a conversation took an average of 72.01 seconds.
We calculated the time-gap between user's first message and workers' first accepted message in each conversational session\footnote{The requester's reputation and workers' trust influence recruiting time. The reported response times in this section only consider the 240 conversations occurred after seven days of our system released, i.e., 2016-05-27 EDT.}.
The first response from workers took 72.01 seconds on average (SD=87.09).
The distribution of the response time of the first crowd message is shown in Figure~\ref{fig:session-response}.
25.00\% of conversations received the first crowd response within 30 seconds,
60.00\% of conversations received the first crowd response within 1 minutes,
and 88.33\% of conversations received the first crowd response with 2 minutes.


In sum, our approach was demonstrated to be able to support a 24-hour on-demand service with a reasonable budget\footnote{We updated \chorus after the submission deadline so that it dynamically decides the number of assignments to post based on the number of workers waiting in the retainer.
It also now sends workers back to the retainer if they did not reach the minimal points in the conversation task.
Furthermore, we adjusted some system parameters such as reward points of each action of workers and the number of workers we recruited for each conversation.
The new system resulted in a much economic price, which is \$2.29 (SD=0.86) per conversational session on average.
As a trade-off, the average response time slightly increased to 110.43 sec (SD=155.84).
}.
We recruited workers by simply posting HITs and turning the untaken assignments into retainers after a conversation is over.
Retainers in our system served as a light-weight traffic buffer to avoid unexpectedly long latency of MTurk.
When a conversational session ends early by incorrect judgement of workers, the retainers can also quickly direct workers to continue with the conversation.
The limitation of this approach is that it heavily relies on the performance of the crowdsourcing platform such as MTurk.
As shown in Figure~\ref{fig:session-response}, several conversations' response time of the first crowd message remain longer than 5 minutes.
We are also aware that the latency of MTurk could be quite long (e.g., 20 to 30 minutes) in some rare occasions.
This suggests that a more sophisticated recruiting model which can adopt to platform's traffic status might be required.

\section{Challenge 4: When Consensus Is Not Enough}

We identified four question types for which workers had difficulty reaching consensus: {\em (i)} questions about the agent's identity and personality, {\em (ii)} subjective questions, {\em (iii)} questions that explicitly referred to workers, and {\em (iv)} requests that asked workers to perform an action.

\subsection{Collective Identity and Personality}

Curious users frequently asked \chorus about its identity, meta data, or personality. The answers to these questions were often inconsistent across sessions run by different group of workers. 
For example, the following user asked where \chorus is located:

\begin{itemize}[label={}]
    \item \textbf{user:} I'm in Pittsburgh. Where are you?
\vspace{-.4pc} \item \textbf{crowd: I'm in the United Kingdom.}
\end{itemize}

\noindent Another user asked \chorus the same question, but got a different answer:

\begin{itemize}[label={}]
    \item \textbf{user:} where are you?
\vspace{-.4pc} \item \textbf{crowd: I am in Florida, where are you}
\end{itemize}

\noindent Sometimes the user asked questions about the agent itself, which the workers did not have an answer for, tending to respond with their personal status, {\em i.e.}, the following example:

\begin{itemize}[label={}]
    \item \textbf{user:} I was wondering about your name. Why is it Chorus Bot?
\vspace{-.4pc} \item \textbf{crowd: I am not sure. I'm new to this.}
\vspace{-.4pc} \item \textbf{user:} How long has it been for you here?
\vspace{-.4pc} \item \textbf{crowd: Is there anything I can help you with?}
\vspace{-.4pc} \item \textbf{crowd: About 3 minutes}
\end{itemize}

\subsection{Subjective Questions}

Users also asked subjective questions, which workers often could not agree on.
As a consequence, users would get a set of answers that obviously came from different people. 
The following example is a question about religion:

\begin{itemize}[label={}]
    \item \textbf{user:} Do you believe Bible is God's word?
\vspace{-.4pc} \item \textbf{crowd: Is that all?}
\vspace{-.4pc} \item \textbf{crowd: Evolution can't be disproven, but neither can creationism in a sense.}
\vspace{-.4pc} \item{[Few messages later.]}
\vspace{-.4pc} \item \textbf{crowd: This worker's opinion is that God does not exist.}
\vspace{-.4pc} \item \textbf{crowd: I believe in a God, but not necessarily all of the things in the Bible}
\end{itemize}

\noindent One user also asked questions about politics:

\begin{itemize}[label={}]
    \item \textbf{user:} who should be the democratic nominee for the presidential race? 
\vspace{-.4pc} \item \textbf{crowd: Bernie Sanders, obviously.}
\vspace{-.4pc} \item \textbf{crowd: Bernie!}
\vspace{-.4pc} \item \textbf{crowd: Hillary Clinton}
\end{itemize}

\subsection{Explicit Reference to Workers}

Curious users also asked explicit questions about crowd workers, including the source of crowd workers, the platform, the worker interface, or the identity of workers.
The following is a typical example:

\begin{itemize}[label={}]
   \item \textbf{user:} who's actually answering these questions
\vspace{-.4pc} \item \textbf{crowd: It's actually a group of workers.}
\vspace{-.4pc} \item \textbf{crowd: A Crowd Worker}
\vspace{-.4pc} \item \textbf{user:} who's in the crowd
\vspace{-.4pc} \item \textbf{crowd: People who have exceptional internet skills.}
\end{itemize}

\noindent Sometimes workers also spontaneously identified themselves and explained their status to the user, which broke the illusion of Chorus being a single agent:

\begin{itemize}[label={}]
   \item \textbf{user:} How come your English is so bad ?
\vspace{-.4pc} \item{[Workers apologize. One worker said ``English is my secondary language''.]}
\vspace{-.4pc} \item \textbf{user:} what's your first language ?
\vspace{-.4pc} \item \textbf{crowd: Crowd 43 - first language is Malayalam}
\vspace{-.4pc} \item \textbf{crowd: There are several of us here my first language is English May I help you find a good place to eat in Seattle?}
\vspace{-.4pc} \item \textbf{crowd: I am worker 43, so you wrote to me or to some one else?}
\end{itemize}

\subsection{Requests for Action}

Some users asked \chorus to perform tasks for them,
such as booking a flight, reserving a restaurant, or making a phone call.
In the following conversation, workers agreed to reserve tables in a restaurant for the user:

\begin{itemize}[label={}]
    \item {[Workers suggested the user to call a restaurant's number to make a reservation.]}
\vspace{-.4pc} \item \textbf{user:} Chorus Bot can't reserve tables :( ?
\vspace{-.4pc} \item \textbf{crowd: I can reserve a table for you if you prefer}
\vspace{-.4pc} \item \textbf{crowd: what time and how many people?}
\end{itemize}

\noindent We were interested to see that workers often agreed to perform small tasks,
but users rarely provided the necessary information for them to do so. We believe these users were likely only exploring what Chorus could do.

\section{Discussion}

During our Chorus deployment, we encountered a number of challenges, including difficulty in finding boundaries between tasks, protecting workers from malicious users, scaling worker recruiting models to mid-sized deployments, and maintaining collective identity over multiple dialog turns. All represent future challenges for research in this area.

\paragraph{Qualitative Feedback}
During the study, we received many emails from both workers and users on a daily basis.
They gave us a lot of valuable feedback on the usage and designs of the system.
We also directly communicated with workers via \chorus by explicitly telling workers ``I am the requester of this HIT'' and asking for feedback.
In general, workers are curious about the project, and several people contacted us just for more details. For instance, workers asked where users were coming from and wondered if it was always the same person asking the questions.
Workers also wanted to know what information users could see (e.g., one worker asked ``Does a new user sees the blank page or the history too?'' in a \chorus -based conversation with us).
The general feedback we received from emails and MTurk forums (e.g., Turkopticon\footnote{Turkopticon: https://turkopticon.ucsd.edu/}) is that workers overall found our tasks very interesting to complete. 
Users also provided feedback via email.
Many were curious about the intended use of this system.
Some users enjoyed talking with \chorus and were excited that the system actually understood them.

\paragraph{How did users use \chorus?}
When users asked us how should they use \chorus, we told them we do not really know, and encouraged them to explore all possibilities.
Interestingly, users used \chorus in a range of unexpected ways:
some users found it very helpful for brainstorming or collecting ideas (e.g., gift ideas for the user's daughter);
one user asked crowd workers to proofread a paragraph and told us it actually helped;
one user tried to learn Spanish from a worker who happened to be a native speaker.
Members of our research group even tried to use \chorus to help collect literature related to their research topics and actually cited a few of them in the paper.
We also observed that several users discussed their personal problems such as relationship-related issues.
These uses of \chorus are all very creative, and beyond what was initially anticipated either by this work or by prior work.
We are looking forward to seeing additional creative usages of \chorus in future deployment.

\section{Conclusion}



In this paper, we have described our experience deploying Chorus with real users.
We encountered a number of problems during our deployment that did not come about in prior lab-based research studies of crowd-powered systems, which will be necessary to make a large-scale deployment of Chorus feasible. We believe many of these challenges likely generalize to other crowd-powered systems, and thus represent a rich source of problems for future research to address.


 


\section{Acknowledgements}
This research was supported by the Yahoo! InMind Project~\cite{inmindVision16} and the National Science Foundation (\#IIS-1149709).
We thank Saiph Savage for the discussion and help.
We also thank the workers on Mechanical Turk who operated Chorus.

\bibliographystyle{aaai}

\begin{thebibliography}{}

\vspace{-.05pc}\bibitem[\protect\citeauthoryear{Allen \bgroup et al\mbox.\egroup
  }{2001}]{allen2001toward}
Allen, J.~F.; Byron, D.~K.; Dzikovska, M.; Ferguson, G.; Galescu, L.; and
  Stent, A.
\newblock 2001.
\newblock Toward conversational human-computer interaction.
\newblock {\em AI magazine} 22(4):27.

\vspace{-.04pc}\bibitem[\protect\citeauthoryear{Azaria and Hong}{2016}]{inmindVision16}
Azaria, A., and Hong, J.
\newblock 2016.
\newblock Recommender system with personality.
\newblock In {\em ACM RecSys 2016}.

\vspace{-.04pc}\bibitem[\protect\citeauthoryear{Baron}{2010}]{baron2010discourse}
Baron, N.~S.
\newblock 2010.
\newblock Discourse structures in instant messaging: The case of utterance
  breaks.
\newblock {\em Language@Internet} 7(4):1--32.

\vspace{-.04pc}\bibitem[\protect\citeauthoryear{Bernstein \bgroup et al\mbox.\egroup
  }{2011}]{bernstein2011crowds}
Bernstein, M.~S.; Brandt, J.; Miller, R.~C.; and Karger, D.~R.
\newblock 2011.
\newblock Crowds in two seconds: Enabling realtime crowd-powered interfaces.
\newblock In {\em UIST 2011},  33--42.

\vspace{-.08pc}\bibitem[\protect\citeauthoryear{Bernstein \bgroup et al\mbox.\egroup
  }{2015}]{bernstein2015soylent}
Bernstein, M.~S.; Little, G.; Miller, R.~C.; Hartmann, B.; Ackerman, M.~S.;
  Karger, D.~R.; Crowell, D.; and Panovich, K.
\newblock 2015.
\newblock Soylent: a word processor with a crowd inside.
\newblock {\em Communications of the ACM} 58(8):85--94.

\vspace{-.08pc}\bibitem[\protect\citeauthoryear{Bigham \bgroup et al\mbox.\egroup
  }{2010}]{VizWiz2010}
Bigham, J.~P.; Jayant, C.; Ji, H.; Little, G.; Miller, A.; Miller, R.~C.;
  Miller, R.; Tatarowicz, A.; White, B.; White, S.; et~al.
\newblock 2010.
\newblock Vizwiz: nearly real-time answers to visual questions.
\newblock In {\em UIST 2010},  333--342.

\vspace{-.08pc}\bibitem[\protect\citeauthoryear{Bohus and Rudnicky}{2009}]{bohus2009ravenclaw}
Bohus, D., and Rudnicky, A.~I.
\newblock 2009.
\newblock The ravenclaw dialog management framework: Architecture and systems.
\newblock {\em Computer Speech \& Language} 23(3):332--361.

\vspace{-.08pc}\bibitem[\protect\citeauthoryear{Chai \bgroup et al\mbox.\egroup
  }{2002}]{chai2002natural}
Chai, J.; Horvath, V.; Nicolov, N.; Stys, M.; Kambhatla, N.; Zadrozny, W.; and
  Melville, P.
\newblock 2002.
\newblock Natural language assistant: A dialog system for online product
  recommendation.
\newblock {\em AI Magazine} 23(2):63.

\vspace{-.08pc}\bibitem[\protect\citeauthoryear{Difallah, Demartini, and
  Cudr{\'e}-Mauroux}{2012}]{difallah2012mechanical}
Difallah, D.~E.; Demartini, G.; and Cudr{\'e}-Mauroux, P.
\newblock 2012.
\newblock Mechanical cheat: Spamming schemes and adversarial techniques on
  crowdsourcing platforms.
\newblock In {\em CrowdSearch},  26--30.

\vspace{-.08pc}\bibitem[\protect\citeauthoryear{Gupta \bgroup et al\mbox.\egroup
  }{2006}]{gupta2006t}
Gupta, N.; Tur, G.; Hakkani-Tur, D.; Bangalore, S.; Riccardi, G.; and Gilbert,
  M.
\newblock 2006.
\newblock The at\&t spoken language understanding system.
\newblock {\em IEEE Audio, Speech, and Language Processing},
  14(1):213--222.

\vspace{-.08pc}\bibitem[\protect\citeauthoryear{Huang, Azaria, and
  Bigham}{2016}]{huang2016instructablecrowd}
Huang, T.-H.~K.; Azaria, A.; and Bigham, J.~P.
\newblock 2016.
\newblock Instructablecrowd: Creating if-then rules via conversations with the
  crowd.
\newblock In {\em CHI 2016 - Extended Abstracts},  1555--1562.

\vspace{-.08pc}\bibitem[\protect\citeauthoryear{Huang, Lasecki, and
  Bigham}{2015}]{huang2015guardian}
Huang, T.-H.~K.; Lasecki, W.~S.; and Bigham, J.~P.
\newblock 2015.
\newblock Guardian: A crowd-powered spoken dialog system for web apis.
\newblock In {\em HCOMP 2015}.

\vspace{-.08pc}\bibitem[\protect\citeauthoryear{Ipeirotis, Provost, and
  Wang}{2010}]{IpeirotisMturkQuality2010}
Ipeirotis, P.~G.; Provost, F.; and Wang, J.
\newblock 2010.
\newblock Quality management on amazon mechanical turk.
\newblock In {\em ACM SIGKDD Workshop on HCOMP},  64--67.

\vspace{-.08pc}\bibitem[\protect\citeauthoryear{Isaacs \bgroup et al\mbox.\egroup
  }{2002}]{isaacs2002character}
Isaacs, E.; Walendowski, A.; Whittaker, S.; Schiano, D.~J.; and Kamm, C.
\newblock 2002.
\newblock The character, functions, and styles of instant messaging in the
  workplace.
\newblock In {\em CSCW 2002},  11--20.

\vspace{-.08pc}\bibitem[\protect\citeauthoryear{Lasecki \bgroup et al\mbox.\egroup
  }{2011}]{lasecki2011real}
Lasecki, W.~S.; Murray, K.~I.; White, S.; Miller, R.~C.; and Bigham, J.~P.
\newblock 2011.
\newblock Real-time crowd control of existing interfaces.
\newblock In {\em UIST 2011},  23--32.

\vspace{-.08pc}\bibitem[\protect\citeauthoryear{Lasecki \bgroup et al\mbox.\egroup
  }{2012}]{lasecki2012real}
Lasecki, W.; Miller, C.; Sadilek, A.; Abumoussa, A.; Borrello, D.; Kushalnagar,
  R.; and Bigham, J.
\newblock 2012.
\newblock Real-time captioning by groups of non-experts.
\newblock In {\em UIST 2012},  23--34.

\vspace{-.08pc}\bibitem[\protect\citeauthoryear{Lasecki \bgroup et al\mbox.\egroup
  }{2013a}]{chorus2}
Lasecki, W.~S.; Thiha, P.; Zhong, Y.; Brady, E.; and Bigham, J.~P.
\newblock 2013a.
\newblock Answering visual questions with conversational crowd assistants.
\newblock In {\em ASSETS 2013}, 18:1--18:8.

\vspace{-.08pc}\bibitem[\protect\citeauthoryear{Lasecki \bgroup et al\mbox.\egroup
  }{2013b}]{Chorus2013}
Lasecki, W.~S.; Wesley, R.; Nichols, J.; Kulkarni, A.; Allen, J.~F.; and
  Bigham, J.~P.
\newblock 2013b.
\newblock Chorus: A crowd-powered conversational assistant.
\newblock In {\em UIST 2013}, UIST '13,  151--162.

\vspace{-.08pc}\bibitem[\protect\citeauthoryear{Lasecki \bgroup et al\mbox.\egroup
  }{2014}]{lasecki2014glance}
Lasecki, W.~S.; Gordon, M.; Koutra, D.; Jung, M.~F.; Dow, S.~P.; and Bigham,
  J.~P.
\newblock 2014.
\newblock Glance: Rapidly coding behavioral video with the crowd.
\newblock In {\em UIST 2014},  551--562.

\vspace{-.08pc}\bibitem[\protect\citeauthoryear{Lasecki, Kamar, and
  Bohus}{2013}]{lasecki2013conversations}
Lasecki, W.~S.; Kamar, E.; and Bohus, D.
\newblock 2013.
\newblock Conversations in the crowd: Collecting data for task-oriented dialog
  learning.
\newblock In {\em HCOMP 2013}.

\vspace{-.08pc}\bibitem[\protect\citeauthoryear{Lasecki, Teevan, and
  Kamar}{2014}]{lasecki2014information}
Lasecki, W.~S.; Teevan, J.; and Kamar, E.
\newblock 2014.
\newblock Information extraction and manipulation threats in crowd-powered
  systems.
\newblock In {\em CSCW 2014},  248--256.

\vspace{-.08pc}\bibitem[\protect\citeauthoryear{Raux and Eskenazi}{2009}]{raux2009finite}
Raux, A., and Eskenazi, M.
\newblock 2009.
\newblock A finite-state turn-taking model for spoken dialog systems.
\newblock In {\em NAACL 2009},  629--637.

\vspace{-.08pc}\bibitem[\protect\citeauthoryear{Salisbury, Stein, and
  Ramchurn}{2015}]{salisbury2015real}
Salisbury, E.; Stein, S.; and Ramchurn, S.
\newblock 2015.
\newblock Real-time opinion aggregation methods for crowd robotics.
\newblock In {\em AAMAS 2015},  841--849.

\vspace{-.08pc}\bibitem[\protect\citeauthoryear{Savenkov, Weitzner, and
  Agichtein}{2016}]{savenkovcrowdsourcing}
Savenkov, D.; Weitzner, S.; and Agichtein, E.
\newblock 2016.
\newblock Crowdsourcing for (almost) real-time question answering.
\newblock In {\em Workshop on Human-Computer Question
  Answering, NAACL 2016}.

\vspace{-.08pc}\bibitem[\protect\citeauthoryear{Teevan, Dumais, and
  Liebling}{2008}]{JaimeUserIntent2008}
Teevan, J.; Dumais, S.~T.; and Liebling, D.~J.
\newblock 2008.
\newblock To personalize or not to personalize: Modeling queries with variation
  in user intent.
\newblock In {\em SIGIR 2008},
  163--170.

\vspace{-.08pc}\bibitem[\protect\citeauthoryear{Von~Ahn and Dabbish}{2004}]{EspGame2004}
Von~Ahn, L., and Dabbish, L.
\newblock 2004.
\newblock Labeling images with a computer game.
\newblock In {\em CHI 2004},  319--326.

\vspace{-.08pc}\bibitem[\protect\citeauthoryear{Vuurens, de Vries, and
  Eickhoff}{2011}]{vuurens2011much}
Vuurens, J.; de~Vries, A.~P.; and Eickhoff, C.
\newblock 2011.
\newblock How much spam can you take? an analysis of crowdsourcing results to
  increase accuracy.
\newblock In {\em CIR 2011)},  21--26.

\end{thebibliography}
\fontsize{10}{9.7}\selectfont

\end{document}